\documentclass[11pt,onecolumn,amssymb,aps,nofootinbib]{revtex4}
\usepackage{amsmath, amsthm, amscd, amssymb}
\usepackage{graphicx, braket}
\usepackage{bm}
\usepackage{bbm}
\usepackage{epstopdf}

\begin{document}

\title{\bf  Equation of state, and atomic electron effective potential, for a Weyl scaling invariant dark energy}

\bigskip

\author{Stephen L. Adler}
\email{adler@ias.edu} \affiliation{Institute for Advanced Study,
1 Einstein Drive, Princeton, NJ 08540, USA.}

\begin{abstract}
A key attribute of dark energy is the equation of state parameter $w={\rm pressure}/{\rm energy~density}$, and this has been recently measured observationally,  giving values close to $-1$.  In this paper we calculate the $w$ parameter characterizing the novel Weyl scaling invariant dark energy that we have analyzed in a series of papers, and show that it is compatible with experiment.  We also derive the atomic electron effective potential induced by dark energy from the electron geodesic equation, which can be applied to the  evaluation of energy level shifts in Rydberg atoms.

\end{abstract}

\vfill\eject

\maketitle
\section{Introdution}

\subsection{Weyl scaling invariant dark energy action}

In  papers over the last nine years    (reviewed in \cite{A1}) we have studied consequences of the postulate that ``dark energy'' arises from an action constructed from
nonderivative metric components so as to be invariant under the Weyl scaling $g_{\mu\nu}(x) \to \lambda(x) g_{\mu\nu}(x)$, where $\lambda(x)$ is a general scalar function.  This novel dark energy action is given by
\begin{equation}\label{dark}
S_{\rm dark~energy}=-\frac{\Lambda}{8 \pi G} \int d^4x  ({}^{(4)}g)^{1/2}(g_{00})^{-2}~~~,
\end{equation}
where $\Lambda$ is the cosmological constant, $G$ is Newton's constant, and ${}^{(4)}g=-\det({g_{\mu\nu})}$.  The action of Eq. \eqref{dark} is compatible with a 3+1 foliation of spacetime associated with the cosmic microwave background rest frame, and is three-space, although not four-space, generally covariant.  Equation \eqref{dark}  reduces to the usually assumed dark energy action in homogeneous cosmological contexts with $g_{00}(x)\equiv 1$, as assumed in the usual $\Lambda CDM$ cosmology. So the successes of the standard $\Lambda CDM$ cosmology are fully consistent with the replacement of the usual dark energy action by Eq. \eqref{dark}  The extra factor $(g_{00})^{-2}$ in Eq. \eqref{dark}  gives rise to novel cosmological effects  only when there are cosmological inhomogeneities, and only in situations where the very small cosmological constant $\Lambda$ becomes relevant.

\subsection{Skeptical questions, motivations, and calculational procedure}

The unconventional nature of the action of Eq. \eqref{dark} has raised many questions, which are addressed in detail in the review \cite{A1} (which is available without subscription as arXiv:2111.12576).  We briefly address a number of them here, and then make related procedural comments.

\begin{itemize}

\item  Doesn't Eq. \eqref{dark} violate the principles of special and general relativity?  As originally formulated, the relativity principle of special relativity  states that an absolute state of uniform motion is not observable. But this is now empirically falsified by observation of a dipole component in the Cosmic Microwave Background (CMB)   radiation, from which one can infer our solar system's state of uniform motion with respect to the rest frame of the CMB, which can be taken as a reference inertial frame.

    A less restrictive form of the relativity principle states that all physical laws should be formulated in a manner not referring to an inertial frame, and this is incorporated into the Einstein-Hilbert (E-H) gravitational action, which is four-space general coordinate invariant, and agrees with all non-cosmological tests of general relativity.  But why should one assume that the very small additional ``dark energy'' action, that describes the late-time accelerated expansion of the universe, obeys the same symmetry principles as the E-H action?

    In the particle physics context, for years it was assumed that since the strong and electromagnetic interactions are governed by a principle of parity invariance, the weak interactions must be also, but this assumption was falsified by experiment.  By analogy in the gravitational context, there is no reason to assume that the dark energy action, the origin of which is quite mysterious,  is governed by the same symmetry principles as the E-H action.  It is possible in principle that although the E-H action is constructed in accordance with the rules of four-space general coordinate invariance,  the very small dark energy action may be constructed in a manner that violates these rules, but is inspired instead by the 3+1 foliation of space-time that is needed to calculate time-dependent spacetime geometries.  The calculations of \cite{previous2} show that because $\Lambda$ is very small, the action of Eq. \eqref{dark} predicts only unobservably small corrections to the standard solar system general relativity tests.  And further calculations in \cite{scalarwave} using the theory of cosmological perturbations show that the residual three-space general coordinate invariance of Eq. \eqref{dark} suffice to prohibit propagating scalar gravitational waves.

\item What about the ``hierarchy'' and ``fine tuning'' problems associated with the cosmological constant?  If $\Lambda$ is interpreted as a vacuum energy density,
    \begin{equation}\label{vacen}
    \rho_{\rm vac}^\Lambda = \frac{\Lambda}{8 \pi G}  \simeq (2 \times 10^{-3} {\rm eV})^4~~~,
    \end{equation}
    it is mysterious why this is so much smaller than typical particle  physics vacuum fluctuation energies of order
    \begin{align}\label{paren}
    & \rho_{\rm vac}^{\rm QM} \sim  M_{\rm Planck}^4 \simeq (10^{28} {\rm eV})^4~~~,\cr
    & \rho_{\rm vac}^\Lambda /\rho_{\rm vac}^{\rm QM}\sim  10^{-120}~~~.\cr
    \end{align}
    This is yet another example of the ``hierarchy problem'' of understanding the smallness of particle masses (including the electron, up and down quark,  and neutrino masses) compared to the Planck mass.  This problem is still present with the assumed action of Eq. \eqref{dark}, since the coefficient $\Lambda$ there is given its observational value.  But if one assumes the conventional dark energy action given by Eq. \eqref{dark} with $g_{00} \equiv 1$,  then there is the additional ``fine-tuning'' problem of understanding why vacuum fluctuation energies of bosons and fermions cancel to 1 part in $10^{120}$, but don't cancel to zero.  If one instead one postulates that so-called ``dark energy'' is {\it not} an energy at all, but instead is governed by a principle of Weyl scaling invariance   (which requires vacuum fluctuation energies to cancel to exactly zero, since the conventional dark energy action is not Weyl scaling invariant), one is led to Eq. \eqref{dark} as an alternative to the conventional interpretation of the cosmological constant as a residual uncancelled vacuum energy, and the fine-tuning problem is eliminated.

    Why  postulate Weyl scaling invariance?  It is suggested by the idea that the underlying theory for both general relativity and particle physics may be scale, and likely also conformal invariant, which leads as a natural postulate to  Weyl scaling invariance of the part of the induced gravitational action that involves no metric derivatives.  Moreover, as Steven Weinberg has often stressed, it is useful in testing current model theories to have alternative ``foils'' that conform to current experiments, but give rise to testable new predictions.   We believe that Eq. \eqref{dark} gives a natural ``foil'' against which to test the conventional assumption that the cosmological constant arises as an energy.

    Finally in this category, we stress that the novel dark energy action is a classical effective action for the classical gravitational metric, as needed for astrophysics and cosmology.  Quantum and pre-quantum degrees of freedom have been integrated out in its derivation, and there are no residual issues of peculiar  quantum field theoretic effects, such as perturbation theory anomalies.

\item  The usual derivation of the conserved stress-energy tensor that is the source term for the Einstein equations makes an assumption of  four-space general coordinate invariance.  So how does one calculate the stress-energy tensor contribution from the novel dark energy action of Eq. \eqref{dark}?  This issue is addressed in the earlier papers reviewed in  \cite{A1}, where a method of covariant completion is proposed.  First one varies the novel dark energy action with respect to the spatial components $g^{ij}$ of the metric, giving  the spatial components $T_{\Lambda \,ij}$ of the stress-energy tensor; then one imposes covariant conservation to compute the space-time components (which vanish) and the time-time component $T_{\Lambda 00}$. In the cosmological context, this procedure is carried out explicitly in
    Eqs. (32)--(35) of  \cite{adlerprd}, for a  line element
    \begin{equation}\label{lineelt}
    ds^2=\alpha^2(t) dt^2-\psi^2(t) d\vec x^2~~~,
    \end{equation}
    with $\psi(0)=0$.
    The results of this calculation  can be summarized as follows,
    \begin{align}\label{stresstensor}
      T_{\Lambda \,ij}=&\frac{\Lambda}{8\pi G} t_{ij}~~~,\cr
      t_{ij}(t)=& -\delta_{ij} \frac{\psi^2(t)}{\alpha^4(t)}      ~~~,\cr
      t_{0i}=& t_{i0} = 0~~~,\cr
      t_{00}(t)=&  3 \frac{\alpha^2(t)}{\psi^3(t)}\int_0^t du \frac{\dot \psi(u) \psi^2(u)}{\alpha^4(u)}~~~,\cr
    \end{align}
    with $\delta_{ij}$ the  Kronecker delta.
Here we have assumed that $t_{00}(0)$ is bounded, which when $\psi(0)=0$ requires taking the lower limit of integration in $t_{00}$ as zero.

\item Just as we have assumed that the E-H gravitational action has the usual form, in the calculations that follow
   we assume that matter actions obey the equivalence principle, with  the flat spacetime Minkowski metric replaced by the gravitational metric $g_{\mu\nu}$.  Then massive and massless particles  follow geodesic trajectories, since as shown by Weinberg \cite{wein},  this is an immediate consequence of the equivalence principle. So our {\it only} non-standard assumption is the choice of the Weyl scaling invariant form of Eq. \eqref{dark} for the dark energy action.

\item
       In two previous papers we have calculated the implications of a scale invariant dark energy for (i) the photon sphere and black hole shadow radii \cite{previous1}, and (ii) light deflection by a central mass, solar system relativity tests, and modifications of the lensing equation \cite{previous2}.  In the subsequent sections of this paper we calculate the dark energy equation of state parameter $w$ as a function of cosmic time, and the dark energy-induced effective potential for an atomic electron.
\end{itemize}

\section{Dark energy equation of state parameter $w$}

From the relativistic perfect fluid form of the stress-energy tensor (with $u_{\mu}$  the fluid four-velocity, $u_0=1$),
\begin{equation}\label{perf}
T_{\Lambda \, \mu \nu}=(p+\rho) u_\mu u_\nu - p g_{\mu \nu},
\end{equation}
compared with  Eq. \eqref{stresstensor}, we can read off the dark energy pressure $p(t)$ and energy density $\rho(t)$, as
\begin{align}\label{prho}
p(t)=&-\frac{1}{\alpha^4(t)}~~~,\cr
\rho(t)=&\frac{3}{\psi^3(t)} \int_0^t du \frac{\dot \psi(u)\psi^2(u)}{\alpha^4(u)}~~~,\cr
\end{align}
with $\dot \psi(u)= \frac{d \psi(u)}{du}$.
We define the equation of state parameter $w(t)$ by $p(t)=w(t) \rho(t)$, and write $\alpha(t)=1+\Phi(t)$, with $\Phi((t)$ the first order scalar perturbation calculated in \cite{adlerprd} and \cite{adlerijmpd} from the modified perturbation equation implied by the dark energy action of Eq. \eqref{dark}.  Working to first order in the small quantity $\Phi( t )$, we get
\begin{align}\label{expansion}
w^{-1}(t)=&\frac{\rho(t)}{p(t)}= -\frac{3}{\psi^3(t)}\int_0^t du \dot \psi(u)\psi^2(u)[1+4\big(\Phi(t)-\Phi(u)\big)]  \cr
=& -1 +D(t) ~~~,\cr
\end{align}
with $D(t)$ given by
\begin{equation}\label{ddef}
D(t)= -\frac{12}{\psi^3(t)} \int_0^t du \dot \psi(u)   \psi^2(u) [\Phi(t)-\Phi(u)]~~~.
\end{equation}
Inverting to find $w(t)$, we get the result
\begin{equation}\label{wresult}
w(t)= -1-D(t)~~~.
\end{equation}
Since $\Phi(t)$ is first order small, we see that the zeroth order approximation to $w(t)$ is just $-1$, with a first order correction given by $-D(t)$.

It is now convenient to change to the dimensionless time variable $x$ used in \cite{adlerprd} and \cite{adlerijmpd},
\begin{equation}\label{xdef}
x=\frac{3}{2} \surd{\Omega_{\Lambda}}H_0^{\rm Pl} t~~~,
\end{equation}
in terms of which Eq. \eqref{ddef} becomes
\begin{equation}\label{ddef1}
D(x)= -\frac{12}{\psi^3(x)} \int_0^x du \frac{d \psi(u)}{du} \psi^2(u) [\Phi(x)-\Phi(u)]~~~.
\end{equation}
Here $H_0^{\rm Pl}\simeq 67 {\rm km} \,{\rm s}^{-1}\, {\rm Mpc}^{-1}$ is the Hubble constant measured by Planck, and the present era $t=t_0$ is defined by $\psi(t_0)=1$. Using the dark energy fraction $\Omega_{\Lambda}=0.679$    and the matter fraction  $\Omega_m=1-\Omega_{\Lambda}=0.321$  of the closure density, the dimensionless time variable  takes the value $x=x_0=1.169$  at the present era.  In evaluating  Eq. \eqref{ddef1} numerically it suffices to use the zeroth order expression for $\psi(x)$ and the expansion of the first order perturbation $\Phi(x)$ in powers of $x$ given in \cite{adlerprd} and \cite{adlerijmpd},
\begin{align}\label{psiandphi}
\psi(x)=&\left(\frac{\Omega_m}{\Omega_{\Lambda}}\right)^{1/3} \left(\sinh(x)\right)^{2/3}~~~,\cr
\Phi(x)=&\Phi(0)[1+\hat{C} x^2 +\hat{D} x^4+O(x^6)]~~~,\cr
\hat{C}=&\frac{2}{11} ~~~, ~~~\hat{D}= -\frac{2}{561}~~~.\cr
\end{align}
The initial value $\Phi(0)\simeq -0.115$ is obtained in \cite{adlerprd} and \cite{adlerijmpd} by using the scale invariant cosmological constant action to fit the observed extra late time acceleration in the expansion of the universe known as the  ``Hubble tension''.  The result of a numerical calculation of $D(x)$,  using  Eqs. \eqref{ddef1} and \eqref{psiandphi} as inputs is given in Fig. 1.  The redshift $z$ corresponding to $x$ can be calculated from the equation $z=\psi^{-1}(x) -1$, and is plotted in Fig. 2.  The $w$ parameter shown in Fig. 1 gives $w(t_0) \simeq -1.05$  at the current epoch  and agrees, to better than two standard deviations, with the small-$z$ value $w=-0.80 \pm 0.18$ obtained recently by  the DES experiment \cite{des}, and the small-$z$ value $w=-1.12 \pm 0.12$ found by eROSITA \cite{rosita}.

\vfill\eject
\begin{figure}[]
\begin{centering}
\includegraphics[scale=0.8]{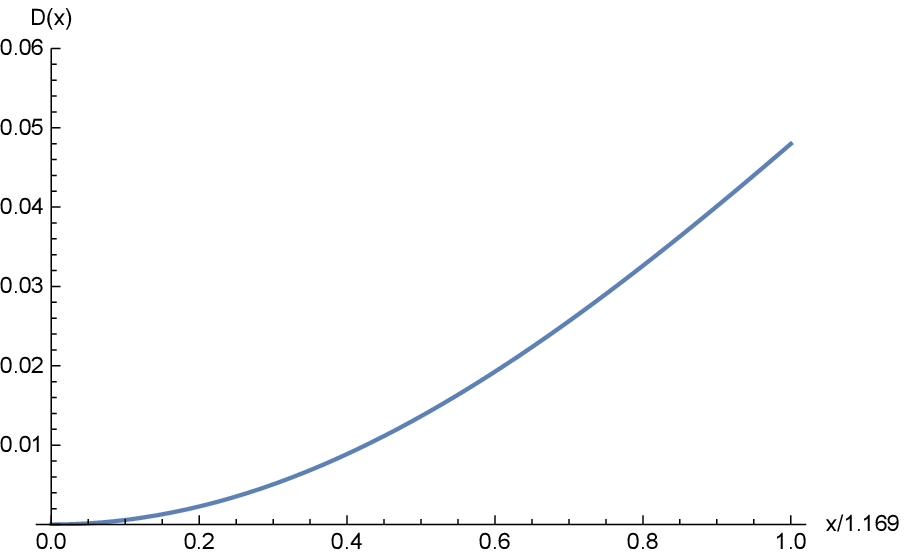}
\caption{ Plot of $D(x)$ versus x/1.169.}
\end{centering}
\end{figure}

\vfill\eject
\begin{figure}[]
\begin{centering}
\includegraphics[scale=0.8]{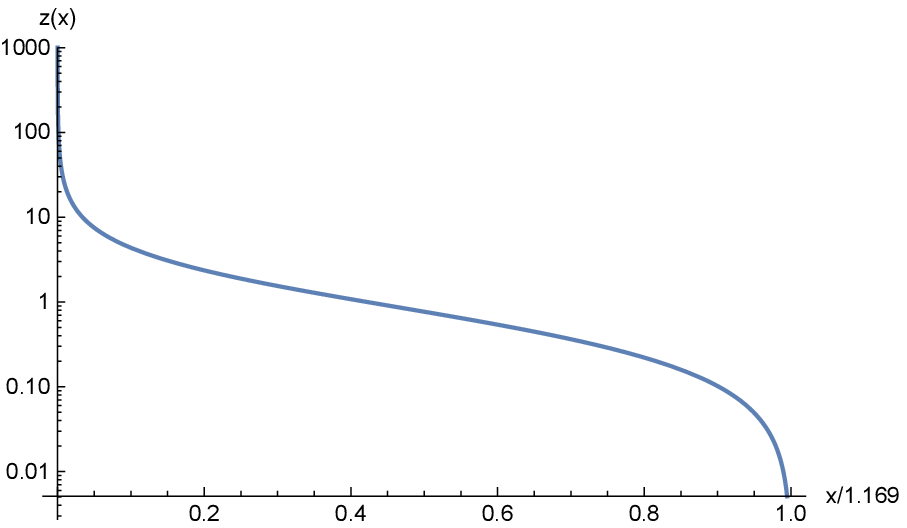}
{zcalc}
\caption{Plot of $z(x)$ versus $x/1.169$. }
\end{centering}
\end{figure}

\section{Electron effective potential from the geodesic equation}

In a paper on Rydberg atom bounds on the cosmological constant \cite{kundu}, Kundu, Pradhan, and Rosenzweig use an effective potential for the electron inferred from an exponentially expanding de Sitter universe.  Our aim in this paper is to show that the same potential can be obtained in a more general way from the geodesic equation for motion of the electron, which gives added insight into how bounds obtained from this potential vary between different models for the
cosmological constant action.

Weinberg in his text``Gravitation and Cosmology'' \cite{wein1}  gives the geodesic equation for a general spherical metric.
For the line element
\begin{equation}\label{linelt1}
ds^2=B(r) dt^2 -A(r) dr^2 -r^2 d\Omega~~~,
\end{equation}
Eq. (8.4.19) of \cite{wein1} gives
\begin{equation}\label{geo}
\frac{A(r)}{B^2(r)}\left(\frac{dr}{dt}\right)^2 +\frac{J^2}{r^2}-\frac{1}{B(r)}=-E~~~,
\end{equation}
where $E_{\rm nr}=(1-E)/2$ plays the role of the non-relativistic energy per unit mass, and $J$ is the angular momentum per unit mass.
Comparing this with the nonrelativistic energy formula
\begin{equation}\label{nrform}
\frac{1}{2} \left(\frac{dr}{dt}\right)^2  +\frac{V_{eff}}{m_e} -E_{nr}=0~~~,
\end{equation}
and doing algebraic rearrangement to eliminate $(dr/dt)^2$ we get a formula for $V_{eff}$,
\begin{equation}\label{Veff}
\frac{V_{eff}}{m_e}=\frac{B^2(r)}{2A(r)} \frac{J^2}{r^2} + \frac{B(r)}{2A(r)}[B(r)-1]- E_{\rm nr}\left[\frac{B^2(r)}{A(r)}-1\right]~~~.
\end{equation}
Since $B(r)$ and $A(r)$ are unity up to small corrections \big(of order the electrostatic and gravitational potentials relative to $m_e$\big), and the dimensionless energy parameter $E_{\rm nr}$ is also very small,    the leading cosmological
constant contribution to the effective potential is given by
\begin{equation}\label{Veffcosm}
V_{eff}\simeq \frac{m_e}{2} [\big(B(r)-1\big)_{\rm cosm}]~~~,
\end{equation}
with  $\big(B(r)-1\big)_{\rm cosm}$ the cosmological constant contribution to $B(r)-1$.
As summarized in \cite{previous2}, one can write for a central mass $M$ in geometrized units,
\begin{align}\label{paramdef1}
A(r)=&1+2M/r-C_A \Lambda r^2 +D_A \Lambda M r+...,\cr
B(r)=&1-2M/r-C_B \Lambda r^2 +D_B \Lambda M r+...,\cr
\end{align}
identifying $\big(B(r)-1\big)_{\rm cosm}=-C_B \Lambda r^2 $,
where the parameters $C_{A,B}$ and $D_{A,B}$ are given in Table I  for a standard dark energy action and for a Weyl scaling invariant \cite{A1} dark energy action.
\begin{table} [ht]
\caption{Parameters $C_A,\,C_B,\,D_A,\,D_B$ for the spherically symmetric line element arising from the conventional and the Weyl scaling invariant dark energy actions \cite{previous2}.}
\centering
\begin{tabular}{c   c c c c}
\hline\hline
dark energy type& $~~C_A~~$  & $~~C_B~~$  &$~~D_A~~$  &  $~~D_B~~$  \\
\hline
~~~~~conventional ~~~~~~~~~~&   -1/3  & 1/3  & 4/3    &  0  \\
Weyl scaling invariant &    1   & 1   &   -10   &  -14 \\
\hline\hline
\end{tabular}
\label{tab1}
\end{table}
So the general formula for the cosmological constant contribution to $V_{eff}$ is
\begin{equation}\label{gen}
V_{eff}=-\frac{m_e \,C_B}{2} \Lambda r^2
\end{equation}
which for the standard dark energy action gives
$V_{eff}=-(1/6) m_e \Lambda r^2$, in agreement with Eq. (7) of \cite{kundu}.  For a Weyl scaling invariant dark energy action, one instead gets  $V_{eff}=-(1/2) m_e \Lambda r^2$ according to Table I.  Thus qualitatively, there is no distinction between the potentials arising in the two cases, as well as for other models of dark energy that take the form of a gravitational action, which will each have a characteristic value of $C_B$.

One might ask what happens if the center for measuring $\vec r$ is not taken as the center of the atom, but a displacement $\vec R$ from the atomic center.  Then $r^2$ in Eq. \eqref{gen} is replaced by $\vec R^{\,2} + 2 \vec R \cdot \vec r +r^2$.  The term $\vec R^{\,2} $ is a constant and does not contribute to energy level differences in Rydberg atoms, and the term $2 \vec R \cdot \vec r$ will average to zero over any inversion invariant squared atomic wave function.  So the calculation of energy differences is invariant with respect to the choice of $\vec R$.

\section{Perturbation theory of Rydberg  atoms}

Given $V_{eff}$, the change in the energy level of a Rydberg atom can be calculated from first order perturbation theory,
\begin{equation}\label{pert}
\Delta E_{cosm}= \langle \Psi| V_{eff} |\Psi \rangle = -\frac{m_e \,C_B}{2} \Lambda  \langle \Psi| r^2  |\Psi \rangle  \simeq
 -\frac{m_e \,C_B}{2} \Lambda  R^2~~~,
\end{equation}
where $R$ is the radius of the Rydberg atom orbit.  Thus, from an upper bound  $|\Delta E_{cosm}|<U$ we get a bound on the cosmological constant
\begin{equation}\label{bound}
\Lambda < \frac{2U}{m_e C_B R^2}~~~.
\end{equation}
If  $U$ is a bound on an energy level difference, $R^2$ in Eq. \eqref{bound} will be an orbit radius-squared difference.   In analogy with the formulas of Eqs. (12)-(15) of \cite{kundu}  this can  be turned into a quantitative bound on $\Lambda$.

 As Kundu et al.  have noted, the energy scale of their bound is much smaller than the energy scale of the standard model, already implying very substantial cancellations if the cosmological constant is interpreted as a vacuum energy. As shown above using a geodesic equation derivation, a  very similar bound is obtained in a dark energy model \cite{A1} in which the cosmological constant does not arise as a vacuum energy. Therefore,  as long as $C_B$ is of order unity, any dark energy action coupled to the gravitational metric gives a bound similar in magnitude to that given by the conventional cosmological constant action.

\section{Acknowledgement}

 I wish to thank Suman Kumar Kundu for calling  my attention to \cite{kundu} and for an informative email correspondence.

Data availability statement:   Data sharing not applicable to this article as no experimental datasets were generated or analysed during the current study.

Keywords:  cosmological constant; dark energy; Weyl scaling invariant; w parameter; electron effective potential; Rydberg atoms

\end{document}